\documentclass[trackchanges,twocolumn]{aastex7}

 \usepackage{amsmath}
     \usepackage{xcolor}
    \usepackage{soul}

\begin{document}

\title{Negative superhumps in cataclysmic variables driven by retrograde apsidal disk precession}

\author[0000-0003-4497-2680]{David Vallet}
\affiliation{Department of Physics and Astronomy, University of Nevada, Las Vegas, 4505 South Maryland Parkway, Las Vegas, NV 89154, USA}
\affiliation{Nevada Center for Astrophysics, University of Nevada, Las Vegas, 4505 South Maryland Parkway, Las Vegas, NV 89154, USA}
\email[]{david.vallet@unlv.edu}  

\author[0000-0003-2401-7168]{Rebecca G. Martin}
\affiliation{Department of Physics and Astronomy, University of Nevada, Las Vegas, 4505 South Maryland Parkway, Las Vegas, NV 89154, USA}
\affiliation{Nevada Center for Astrophysics, University of Nevada, Las Vegas, 4505 South Maryland Parkway, Las Vegas, NV 89154, USA}
\email{}

\author[0000-0002-4636-7348]{Stephen H. Lubow}
\affiliation{Space Telescope Science Institute, 3700 San Martin Drive, Baltimore, MD 21218, USA}
\email{}

\author[0000-0003-2270-1310]{Stephen Lepp}
\affiliation{Department of Physics and Astronomy, University of Nevada, Las Vegas, 4505 South Maryland Parkway, Las Vegas, NV 89154, USA}
\affiliation{Nevada Center for Astrophysics, University of Nevada, Las Vegas, 4505 South Maryland Parkway, Las Vegas, NV 89154, USA}
\email{}

\begin{abstract}
Negative superhumps are photometric modulations in cataclysmic variables with periods slightly shorter than the orbital period. They are usually attributed to retrograde nodal precession of a tilted accretion disk, although the origin and persistence of the tilt remains unexplained. We propose instead that negative superhumps arise from retrograde apsidal precession of an eccentric disk. Using linear eccentric disk theory, we show that the direction of apsidal precession is highly sensitive to disk size and temperature, and that pressure effects can drive retrograde precession even in cool disks. In low mass ratio systems where the 3:1 resonance is within the disk, disk expansion during outbursts may produce opposite precession directions in the inner and outer disk, allowing the temporary coexistence of positive and negative superhumps, and driving dissipation in an extended superoutburst. In higher mass ratio systems where the resonance location is outside of the disk, the resonance width can still extend into the outer parts of the disk, excite eccentricity, and drive apsidal precession. This mechanism explains the prevalence of negative superhumps across a wide range of mass ratios and accretion states, without requiring a long-lived disk tilt. It may also explain how positive superhumps can occur in high mass ratio systems if the disk density builds up in the outer parts of the disk.
\end{abstract}

\keywords{\uat{Cataclysmic variable stars}{203} --- \uat{Semi-detached binary stars}{1443} --- \uat{Accretion}{14}}

\section{Introduction}

Cataclysmic variables are binary star systems in which a white dwarf accretes material from a Roche-lobe filling red dwarf companion. They display a wide range of photometric and spectroscopic variability driven primarily by differences in binary mass ratio and accretion rate. Systems with high mass-transfer rates maintain hot, stable accretion disks and do not show large-amplitude outburst activity. In this category, nova-like systems have high mass ratios while permanent superhumpers have low mass ratios.  Dwarf novae, by contrast, have lower accretion rates and undergo recurrent outbursts caused by the thermal–viscous instability associated with hydrogen ionization in the disk \citep[e.g.][]{Osaki1974,Smak1984,Osaki1996}.
Dwarf novae are further divided into subclasses based on their mass ratios and outburst behavior. SU~Uma stars have low mass ratios and exhibit both normal outbursts and brighter, longer superoutbursts \citep{Warner1995}.  After several normal outbursts, a superoutburst occurs once the disk expands sufficiently \citep{Osaki1989,osaki2003,Osaki2005,Osaki2013,Osaki2014}. 
U Gem and Z Cam systems have higher mass ratios and do not have superoutbursts, only normal outbursts.

Positive superhumps (PSHs) are photometric modulations with periods slightly longer than the orbital period. They are observed in low-mass-ratio systems, including permanent superhumpers (superhumps in these systems are persistent and do not only occur during superoutbursts) and SU~Uma systems during superoutbursts \citep[and sometimes during the immediately preceding normal outburst][]{Osaki2014} as well as in a few high-mass-ratio systems.  PSHs arise from prograde apsidal precession of the eccentric disk, as shown in simulations \citep{Whitehurst1988,Hirose1990,Lubow1991b,Ichikawa1993,Murray1996,Murray1998b,Simpson1998,Smith2007,Oyang2021}. The eccentricity of the disk increases due to a tidal instability involving a mode-coupling mechanism that occurs when the disk extends to the 3:1 resonance  \citep{Lubow1991}. The 3:1 resonance is inside of the disk tidal truncation radius for binary mass ratios $q=M_2/M_1 \le q_{\rm crit} \approx 0.33$ \citep{Hirose1990, Whitehurst1991,Patterson2005}. Pressure introduces a 
contribution to the precession rate that must be included to reproduce observed superhump periods \citep{Lubow1992,Hirose1993,Murray2000,Ogilvie2008}.

Negative superhumps (NSHs) have periods slightly shorter than the orbital period \citep{Sun2024,Sun2024b} and are observed in systems that span a wide range of mass ratios and mass transfer rates \citep[e.g.][]{Kimura2020,Bruch2023}.  In SU~Uma systems, NSHs can persist throughout the superoutburst cycle, although with much smaller amplitudes than PSHs \citep[e.g.][]{Still2010,Ohshima2012,Ohshima2014}. During superoutbursts and the preceding normal outburst, some systems can display both PSHs and NSHs at the same time \citep[e.g.][]{Olech2009}. NSHs are also often observed in high mass ratio binaries, both nova-likes and dwarf novae \citep{Bruch2023}.

NSHs have been attributed to a tilted disk that undergoes retrograde nodal precession \citep{Patterson1993,Smak2009,Kimura2021}. However, no satisfactory tilt mechanism has been proposed. The tilt is subject to strong effects of viscous damping that act
to reduce the tilt on timescales that are too short for NSHs to persist, unless the turbulent viscosity
parameter is much smaller than expected $\alpha \la 10^{-4}$ \citep{King2013}. There is a tidal tilt instability mode-coupling mechanism that acts in analogy with the eccentricity instability mechanism \citep{Lubow1992a}. However, the tilt instability growth rate is only a few percent of the eccentricity growth rate and is then subject to suppression by damping effects.  The tidal tilt instability  was found in hydrodynamic simulations to be too weak to generate tilt and radiation driven warping is also unlikely to provide sufficient torque to account for the tilt \citep{Murray1998}. Hydrodynamic simulations have often imposed a tilt or assumed a tilted magnetic field is able to tilt the disk \citep{Wood2009,Thomas2015}.

In this Letter, we propose that NSHs in cataclysmic variables are driven by retrograde {\it apsidal} precession of an eccentric disk.  Retrograde apsidal precession has been found in many 2D simulations \citep[e.g.][]{Kley2008,Jordan2021,Jordan2024,Ohana2025}, but has not previously explored as a source of NSHs.  In Section~\ref{sec:1d} we describe the 2D and 3D linear equations that describe the evolution of an eccentric disk. In Section~\ref{results}, we consider low mass ratio systems and show how the direction of apsidal precession is sensitive to the outer truncation radius of the disk and the aspect ratio of the disk. We explore how this can explain NSHs in SU~Uma systems. In Section~\ref{highmass}, we show that eccentricity growth and retrograde apsidal precession may be driven even in large mass ratio systems. We draw our conclusions in Section~\ref{sec:concs}.

%\newpage     % temporary seperation
% Highlights are for my own attention 

\section{Eccentric disk model}  \label{sec:1d}

We consider the evolution of an accretion disk around a primary star, a white dwarf,  with mass $M_1$. The companion has mass $M_2$ and the binary has a mass ratio of $q=M_2/M_1$ and separation $a_{\rm b}$. The orbital period of the binary is $P_{\rm orb}=2 \pi /\Omega_{\rm b}$, where $\Omega_{\rm b}=\sqrt{G (M_1+M_2)/a_{\rm b}^3}$. Material in the disk orbits about the primary star at the Keplerian frequency   given by $\Omega = \sqrt{ G M_1 /r^3}$. The disk extends from inner radius $r_{\rm in}=0.01\,a_{\rm b}$ to outer radius $r_{\rm out}$ with surface density $\Sigma\propto R^{-1/2}$, which is assumed to be fixed in time. We explore values for the outer radius in the range $r_{\rm out}=0.42-0.58\, a_{\rm b}$.  The pressure is given by $P = c_{\rm s}^2 \Sigma$, where the sound speed  is given by $c_{\rm s}=(H/r) r \Omega$, and $H/r$ is the aspect ratio of the disk that is taken to be constant with radius.

We analyze the radial distribution, growth, and precession of eccentricity in accretion disks by excitation of the 3:1 Lindblad resonance \citep{GO06,Lubow2010}.
The complex eccentricity of the disk is defined as 
\begin{equation}
E=E(r,t)=e(r,t)\exp{ \left[ i \varpi(r,t)  \right]}
\end{equation}
for the real eccentricity $e$ and the periapse angle $\varpi$, both as a function of radius $r$ and time $t$. For small gradients in $e$ and $\varpi$ the orbits in the disk are nested Keplerian orbits \citep{Ogilvie2001}.

The equation governing the eccentricity evolution of an adiabatic disk is
\begin{equation}
J\frac{\partial E}{\partial t} = i \frac{\partial }{\partial r}\left(a \frac{\partial E}{\partial r} \right) + ibE + JsE. \label{eq:EQ1}
\end{equation}
The functions  $a(r)$, $b(r)$ and $s(r)$ are defined separately for 2D and 3D disks in the following subsections.
The angular momentum per unit radius divided by $\pi$ is given by
\begin{align}
J = 2 r^3 \Omega \Sigma.
\end{align}
We consider both 2D and 3D versions of these equations, which we describe in the next two subsections.

\subsection{2D equations}

In the 2D equations, the functions $a$ and $b$  are given by
\begin{equation}
 a = (\gamma - i\alpha) P r^3
 \end{equation}
 and
 \begin{equation}
 b  = \frac{dP}{dr}r^2 + J\dot{\varpi}_g
\end{equation}
\citep{GO06},
where $\gamma = 3/5$ is the gas adiabatic index, $\alpha=0.1$  is an eccentricity damping parameter related to the bulk viscosity of the disk (rather than the usual shear viscosity parameter of \cite{Shakura1973}).
The gravitational precession rate of a free particle on an eccentric orbit is defined as
\begin{align}
    \dot{\varpi}_g = \frac{1}{4} q \left( \frac{r}{a_{\rm b}} \right)^2 \Omega \, b^{(1)}_{3/2}{ \left(\frac{r}{a_{\rm b}} \right) },
\end{align}
where 
\begin{align}
b^{(1)}_{3/2}{ \left(\frac{r}{a_{\rm b}} \right) } =  \frac{1}{\pi}\int^{2\pi}_0 \frac{ \cos{( x)} ~dx}{ \left( 1+ (r/a_{\rm b})^2 -2(r/a_{\rm b})\cos{x}   \right)^{3/2}   }
\end{align}
is the Laplace coefficient.

\subsection{3D equations}

In the 3D case, the eccentric disc is not in vertical hydrostatic balance because fluid elements undergo changing vertical gravity in time, owing to radial excursions.  The vertical motions couple to horizontal motions that in turn lead to a prograde contribution to the apsidal precession rate involving pressure \citep{Ogilvie2008}. 

For the 3D case, we instead have
\begin{align}
    a = 
    \left(2 - \frac{1}{\gamma} -i\alpha \right)P r^3
\end{align}
and
\begin{align}
    b = \left(4 - \frac{3}{\gamma}\right) r^2 \frac{dP}{dr} + 3\left(1 + \frac{1}{\gamma}\right)Pr + J\dot{\varpi}_g
\end{align}
\citep[e.g.][]{Ogilvie2014,Teyssandier2016}.

\subsection{3:1 resonance forcing term}

The function $s$ in equation~(\ref{eq:EQ1}) is the forcing term acting on the resonance location and is given by
\begin{align}
    s = \chi ~\frac{ \exp{[ -(r-r_{\rm res})^2/w_{\rm res}^2  ]} }{ \sqrt{\pi}w_{\rm res} }   
\label{eq:forcing}
\end{align}
\citep{Lubow2010}\footnote{Note that there is typo in equation 9 in \cite{Lubow2010} that is corrected in an erratum \cite{Lubow2012}.}, where $\chi = 2.08 q^2 \Omega_b r_{\rm res}$ \citep{Lubow1991, Ogilvie07}.
The 3:1 resonance radius is 
\begin{equation}
r_{\rm res} = 3^{-2/3}(1+q)^{-1/3} a_{\rm b}
\label{rres}
\end{equation}
and the resonance width due to gas pressure is 
\begin{equation}
    w_{\rm res} = (H/r)^{2/3}r_{\rm res}
\end{equation}  \citep{Meyer87}.

\subsection{Boundary and initial conditions}

A Neumann boundary condition is applied to both the inner and outer edges of the disk so that
\begin{equation}
 \left.   \frac{dE}{dr}\right|_{r_{\rm in}} = \left. \frac{dE}{dr}\right|_{r_{\rm out}} = 0.
\end{equation}
The initial condition satisfies a normalized eccentricity at the outer edge of the disk with no eccentricity at the inner edge given by
\begin{equation}
   E(r,0) = \cos{\left[ \frac{\pi}{2} \left( \frac{r_{\rm out} - r}{r_{\rm out} - r_{\rm in}}  \right) \right]}.
\end{equation}

\subsection{Normalized Eccentricity Profiles}

We follow similar methods to \cite{Lubow2010} to find the eccentricity profile. Equation~(\ref{eq:EQ1}) is initially solved with Matlab's PDEPE solver based on the methods presented in \cite{Skeel90}. Over time, the disk approaches an exponentially growing eigenmode
in which the eccentricity profile 
with radius normalized by the eccentricity at the outer radius approaches a fixed function. In addition, the disk undergoes apsidal precession at a constant rate.

For this eigenmode, the time dependence of the eccentricity is assumed to have the form $E\propto \exp (i \omega t)$. We can calculate the eccentricity  growth  rate and the precession rate with 
\begin{equation}
i \omega=    \frac{\dot E}{E}=\frac{\dot e}{e}+i\dot \varpi.
\end{equation}
The eccentricity growth rate is $-\Im (\omega)$ and the precession rate is $\Re (\omega)$.
Once the disk reaches an eigenmode, these quantities do not change with radius or time.

We first estimate the complex eigenvalue $\omega$ from the gradient of $E$ calculated using the results of the time-dependent PDEPE solver after a time of 128 orbital periods. We then use this time-dependent result for $\omega$ as an initial estimate to solve the time-independent second order spatial equation~(\ref{eq:EQ1}), which we solve with a Runge-Kutta 4-5 predictor-corrector algorithm. The time independent equation determines $\omega$ by converging on the inner boundary Neumann condition via the shooting method. The outer boundary also has a Neumann condition, while the eccentricity is normalized to $E =1.0$. As the complex eigenvalue $\omega$ has both a real and an imaginary part, we use the Neader Simplex optimization routine to converge on the solution. It should be noted that the mode is not unique and is theorized to be the fastest growing mode \citep{Lubow2010}. Once the eigenvalue $\omega$ for the fastest growing mode is found, we iteratively apply the value of $\omega$ obtained in the last iteration
as an initial guess for $\omega$ in the shooting method with small parameter variations of $H/r$ and $r_{\rm out}$. After many iterations, we are able to efficiently
 carry out a broad parameter sweep.
More details can be found in \cite{Lubow2010}.

\section{Low mass ratio binaries}
\label{results}

In this section, we first consider the eigenmode for some standard disk parameters with a mass ratio of  $q=0.1$, typical of an SU~Uma system.  We then explore a wide range of parameters for $H/r$ and the outer truncation radius of the disk, $r_{\rm out}$.

\subsection{Eccentricity profile }

\begin{figure*}
\centering
\includegraphics[width = 0.49\textwidth]{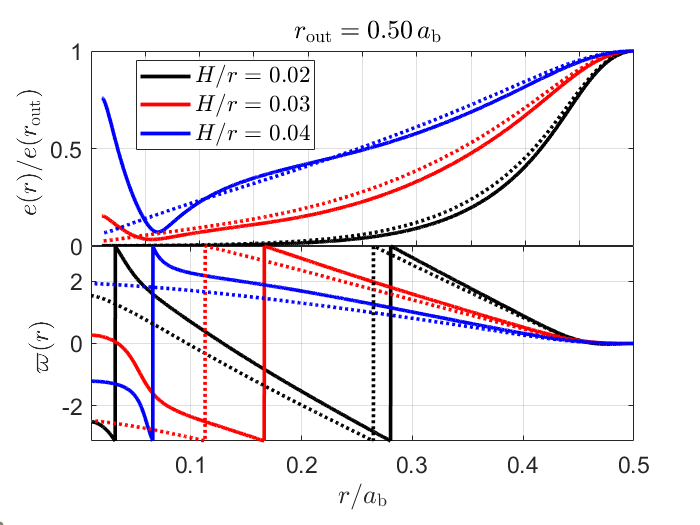} 
\includegraphics[width = 0.49\textwidth]{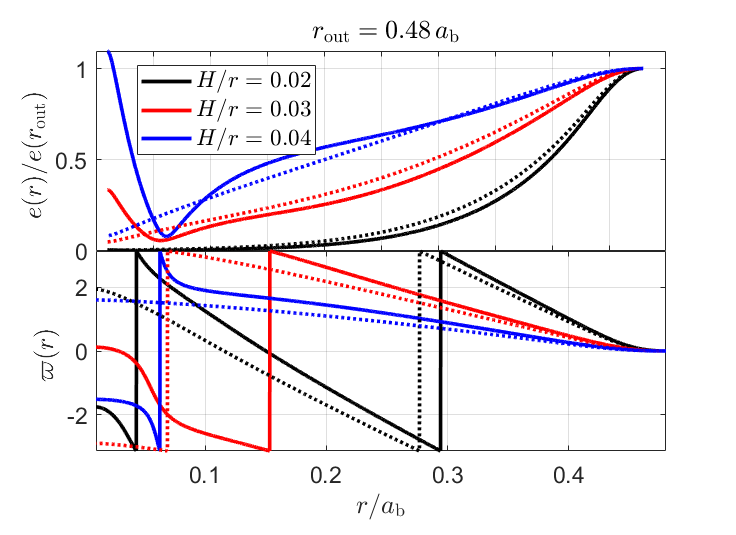} 
\caption{The eccentricity (upper panels) and periapse angle (lower panels) versus radius of the disk for three different values of $H/r$ and $q=0.1$. The outer disk truncation radius is $r_{\rm out}=0.5\,a_{\rm b}$ (left) and  $r_{\rm out}=0.48\,a_{\rm b}$ (right). The dotted lines show the solution to the 2D equations and the solid lines show solution to the 3D equations.   }
\label{fig:delta_HR}
\end{figure*}

Fig.~\ref{fig:delta_HR} shows the eccentricity and the periapse angle as a function of radius for different values of $H/r$ and the outer disk radius is $r_{\rm out}=0.5\,a_{\rm b}$ and $r_{\rm out}=0.48\,a_{\rm b}$. For comparison, the Roche lobe radius is $r_{\rm l}=0.58\,a_{\rm b}$, where
\begin{equation}
    r_{\rm l}=\frac{0.49q'^{2/3}a_{\rm b}}{0.6q'^{2/3}+{\rm ln}(1+q'^{1/3})},
\end{equation}
and $q'=1/q=M_1/M_2$ \citep{Eggleton1983}. Solutions are shown for both the 2D and the 3D equations. The 2D equation solutions are in agreement with \cite{Lubow2010}. The 3D equation solutions are similar except that we see that there can be large eccentricity growth in the innermost parts of the disk
for $H/r \gtrsim 0.02$. In addition, there is more eccentricity growth at the inner disk edge for a smaller outer disk truncation radius.

\begin{figure}
\centering
\includegraphics[width = 0.49\textwidth]{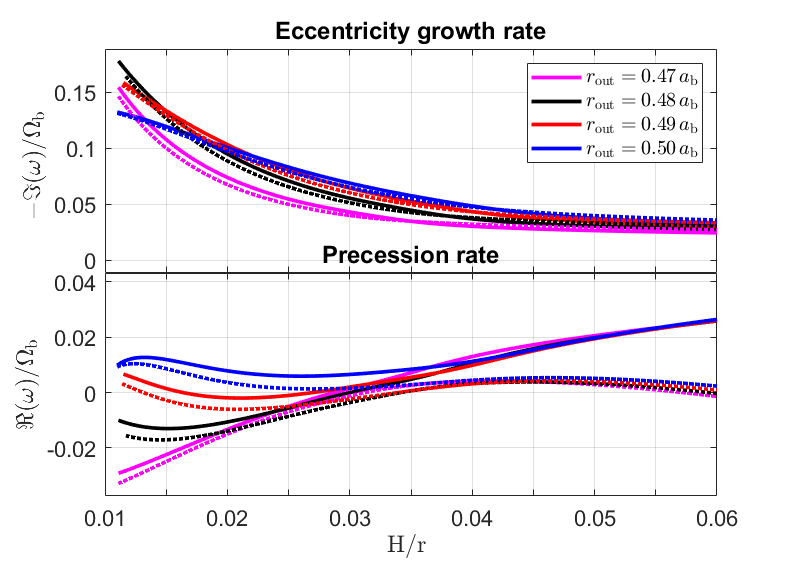} 
\caption{The eccentricity growth rate (upper panel) and precession rate (lower panel) as a function of $H/r$ are shown for various disk outer radii and $q=0.1$. The dotted lines show the solution to the 2D equations and the solid lines show solution to the 3D equations.  }
\label{fig:precession_rate}
\end{figure}

Fig.~\ref{fig:precession_rate} shows the eccentricity growth rate and the apsidal precession rate in the eigenmode as a function of the disk aspect ratio for different values of the outer disk radius. For the eccentricity growth rate, there is not much difference between the 2D and 3D equation solutions. The growth rate decreases with increasing $H/r$. 
Similarly, the growth rate generally decreases with decreasing truncation radius. This is because the width of the resonance moves outside of the truncation radius and so the resonance becomes weaker.

 For low $H/r$ the 2D and 3D equations show similar behavior. However, there are some differences in the apsidal precession rate between the 2D and 3D equations that are seen for larger $H/r$. The apsidal precession rate for the 2D equations turns over and becomes negative for large values of $H/r$  (see also \citet{Kley2008} and \citet{Lubow2010}) while the 3D equations have a precession rate that increases with $H/r$.  The important point to note here is that the apsidal precession can be retrograde for $H/r \lesssim 0.03$, depending on the outer truncation radius of the disk.

For cooler disks (smaller $H/r$), the precession rate decreases with decreasing disk outer radius and  can be negative.
This occurs because the prograde effects of the companion's gravity decrease, while the retrograde effects of pressure increase.
 Surprisingly, at fixed $r_{\rm out} \le 0.48\, a_{\rm b}$ the precession
rate sometimes becomes more retrograde with decreasing $H/r$. The reason is that the precession rate contribution due to pressure also depends on the the square of the eccentricity radial derivative \citep[see equation 36 in][]{GO06} that increases with smaller disk sound speed, as the eccentricity
becomes more narrowly confined to the disk outer edge. Thus, the retrograde effects of pressure can increase with smaller $H/r$.

\begin{figure}
    \centering
    \includegraphics[width=1.0\linewidth]{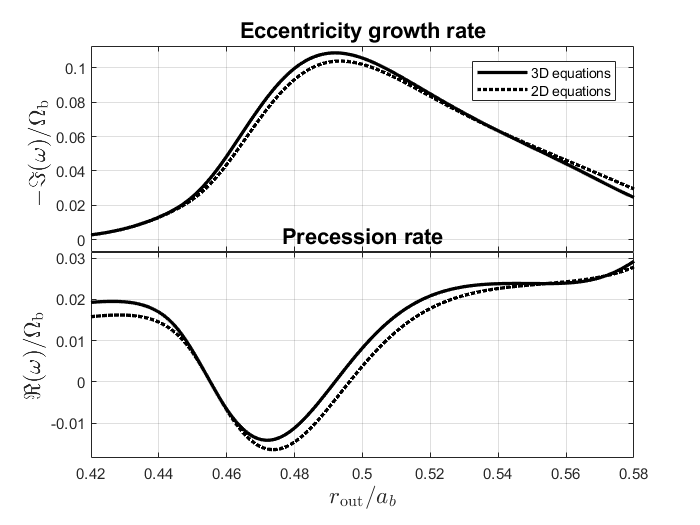}
    \caption{The eccentricity growth rate (upper panel) and precession rate (lower panel) as a function of the disk outer radius, $r_{\rm out}$ with $H/r=0.02$ and $q=0.1$. The dotted lines show the solution to the 2D equations and the solid lines show solution to the 3D equations.}
    \label{fig:rout}
\end{figure}

Given the sensitivity of the precession rate to the outer truncation radius, we now explore a wider range of truncation radii. Fig.~\ref{fig:rout} shows how the eccentricity growth rate and precession rate change with the outer radius for $H/R=0.02$ and $q=0.1$. The resonance radius is $r_{\rm res}=0.466\,a_{\rm b}$ (see equation~\ref{rres}). For small truncation radius $r_{\rm out}\lesssim 0.45\,a_{\rm b}$, the eccentricity growth rate is small. For truncation radius $0.45 \lesssim r_{\rm out}/a_{\rm b} \lesssim 0.49$, the precession is retrograde and the eccentricity growth rate peaks. As the truncation radius increases ($r_{\rm out}\gtrsim 0.49$), the eccentricity growth rate decreases and the precession becomes more prograde.  

\subsection{Negative superhumps during quiescence}

We propose that NSHs in SU~Uma systems occur when the disk becomes eccentric and undergoes retrograde apsidal precession. The disk size is expected to be smaller during quiescence than during an outburst, and this makes retrograde precession more likely (see Fig.~\ref{fig:rout}).

NSHs in SU~Uma systems can persist over several superoutburst timescales \citep{Osaki2013}. If they are driven by apsidal precession of an eccentric disk, the disk must remain somewhat eccentric throughout this time. This can occur either if the disk remains in contact with the resonance, or if the eccentricity damping timescale is long.   During quiescent times, sometimes NSHs are observed but PSHs do not occur. The disk viscosity and temperature are low with typical values being $\alpha=0.01$ and $H/r=0.01$ \citep[e.g.][]{Hameury1998}. Such low values of $H/r$ mean that the inner disk is unlikely to be eccentric if the disk is steady (see Fig.~\ref{fig:delta_HR}), although if the eccentricity damping timescale is long, then the disk may remain eccentric during quiescence.

The communication through the disk may be through viscous communication.  The viscous timescale is
\begin{equation}
    t_{\rm visc}= \frac{1}{\alpha (H/R)^2\Omega}
\end{equation}
\citep{Pringle1981}.
For typical parameters during quiescence and at a radius close to the outer edge of the disk, this is given by
\begin{align}
    t_{\rm visc} \,= \, & 5.6\times 10^4
    \left(\frac{\alpha}{0.01}\right)^{-1} 
    \left(\frac{H/R}{0.01}\right)^{-2} \cr
 &   \times
    \left(\frac{R}{0.5\,a_{\rm b}}\right)^{3/2}  (1+q)^{1/2} P_{\rm orb}.
\end{align}
During quiescence, if $\alpha=0.01$, $H/r=0.01$, and $q=0.1$, then the viscous timescale is $t_{\rm visc}\approx 6 \times 10^4\,P_{\rm orb}$, much longer than the timescale between superoutbursts \citep[which is $\lesssim 2000\, P_{\rm orb}$,][]{Osaki2013}. Therefore, the disk may be able to remain eccentric between outbursts if $\alpha$ and $H/r$ are small enough. Note that the eccentricity decay also depends on the inflowing material, which may help to circularize the disk. Therefore, the eccentricity of the disk during quiescence also depends sensitively upon  the fraction of the disk mass that is accreted during the superoutburst. 

\subsection{Negative superhumps during superoutbursts}

A range of superhump behaviors are observed during superoutbursts, with some systems showing only PSHs, some showing only NSHs, and some showing both simultaneously \citep[e.g.][]{Liu2023,Joshi2025}. During outbursts, typical disk parameters are expected to be $\alpha=0.1-0.3$ and $H/r=0.02-0.03$ \citep{King2013}. With such a large $H/r$, the inner edge of the disk can be eccentric during outbursts (see Fig.~\ref{fig:delta_HR}). Therefore we suggest that NSHs during superoutbursts may be explained by an eccentric inner disk.   NSHs have been observed to be from the inner parts of the disk \citep{Imada2018}. As the disk radius increases during the outburst, at least the outer parts of the disk can undergo prograde precession (see Fig.~\ref{fig:rout}).

It is possible for the apsidal precession to be in different directions at different disk radii, but as the disk eccentricity gets twisted up, this may only be a temporary phenomenon. However, PSHs and NSHs are observed simultaneously only for relatively short periods of time. NSHs are persistent and occur over a timescale of several superorbital periods, while PSHs only occur for brief periods of time, that is, during the superoutbursts and sometimes during the previous normal outburst.

It is also possible that a disk breaks as a result of poor communication and eccentricity growth \citep[e.g.][]{Overton2025}. The disk breaks into disjoint rings, which could make it easier to have apsidal precession that is retrograde in the inner parts of the disk and prograde in the outer parts of the disk. Furthermore, a differentially precessing disk, or even a broken disk, may lead to increased dissipation and, therefore, to more heating, exactly what is observed during superoutbursts. Eccentric disk breaking has been seen before in smoothed particle hydrodynamic simulations \citep{Overton2025}.

The frequency and amplitude of NSHs in SU Uma systems varies over the superoutburst cycle. The frequency is higher during outbursts and decreases during quiescence \citep{Osaki2013}. This may be a result of the changing disk outer radius during the outburst. The amplitude of the NSH decreases as the outburst increases and rebounds at the plateau \citep[see also][]{Sun2024,Sun2024b}. This may be a result of the changing disk eccentricity profile. However, given the time dependent behaviour of a disk undergoing an outburst, understanding these observations in detail requires time-dependent hydrodynamic simulations that we leave to future work.

\subsection{When do negative superhumps appear/disappear?}

Systems may show NSHs for a few superoutburst cycles, and then they can disappear \citep[e.g.][]{Osaki2013}. The appearance/disappearance may be a result of small changes to the disk temperature profile.  A cooler and smaller disk makes retrograde apsidal precession and NSHs more likely (see Figs.~\ref{fig:precession_rate} and~\ref{fig:rout}) and also increases the eccentricity damping timescale. If the disc eccentricity damps quickly in quiescence, then NSHs will not be seen through this mechanism. The presence of NSHs appears to suppress the frequency of dwarf nova outbursts \citep{Ohshima2012}. This could also be the result of a cooler disk that increases the critical surface density required for an outburst, leading to a longer time between outbursts  \citep{Lasota2001}.

\section{High mass ratio binaries}
\label{highmass}

We have seen that NSHs in systems with low mass ratios may be driven by retrograde apsidal precession. We now turn to the question of how superhumps may be driven for higher mass ratio binaries.  In many of these binaries NSHs are observed, however, PSHs typically are not observed.

\begin{figure*}
    \centering
    \includegraphics[width=0.305\linewidth]{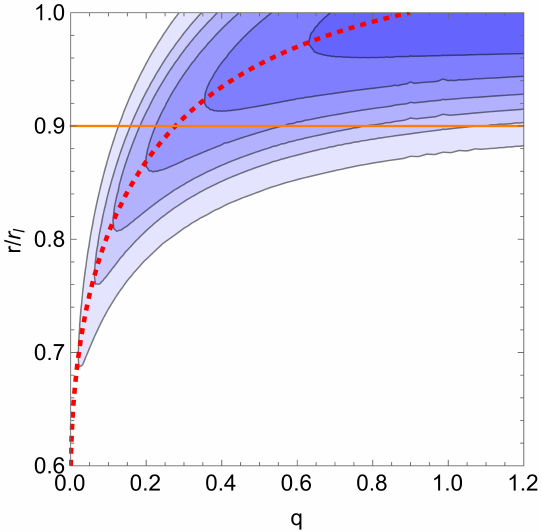}
    \includegraphics[width=0.305\linewidth]{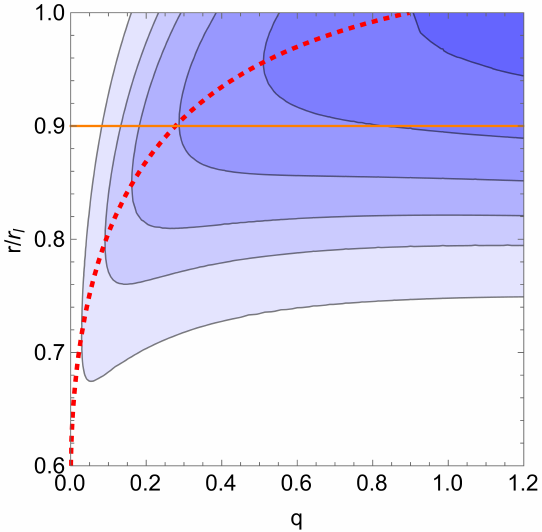}
        \includegraphics[width=0.37\linewidth]{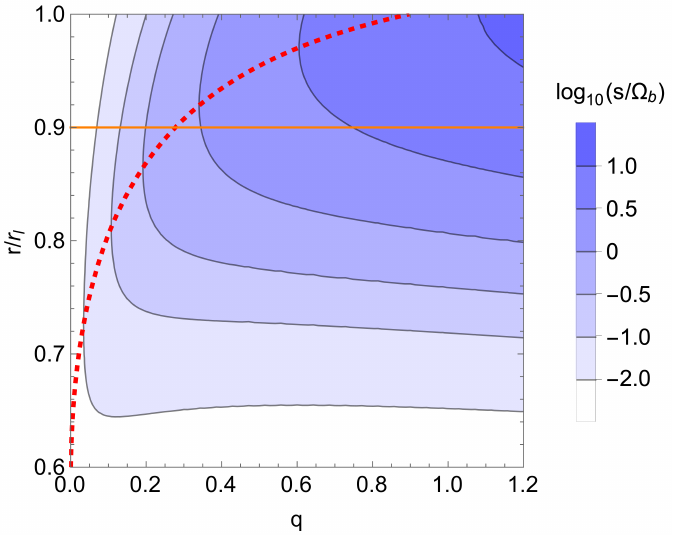}
    \caption{The location of the 3:1 resonance as a function of binary mass ratio (red dashed lines, equation~\ref{rres}). The contours show the forcing term, $s$ (equation~\ref{eq:forcing}), for disk aspect ratios of $H/r=0.01$ (left), 0.03 (middle) and 0.05 (right). The solid orange lines show a typical location of the disk outer radius for a tidally truncated disk at $r_{\rm out}=0.9\,r_{\rm l}$.} 
    \label{fig:width}
\end{figure*}

Fig.~\ref{fig:width} shows the location and strength of the 3:1 resonance as a function of mass ratio. The usual critical mass ratio of $q_{\rm crit} \approx 0.33$ can be found by where the resonance location (red dashed line) crosses the tidal truncation radius of the disk, that is somewhere around $0.85-0.9\, r_{\rm l}$ \citep{Paczynski1977}. The plots show an orange horizontal line at $r_{\rm out}=0.9\,r_{\rm l}$.  The width of the 3:1 resonance can be large enough for the resonance to extend into the outer parts of the disk, even for large binary mass ratios, depending upon $H/r$.

\begin{figure}
    \centering
    \includegraphics[width=\linewidth]{ 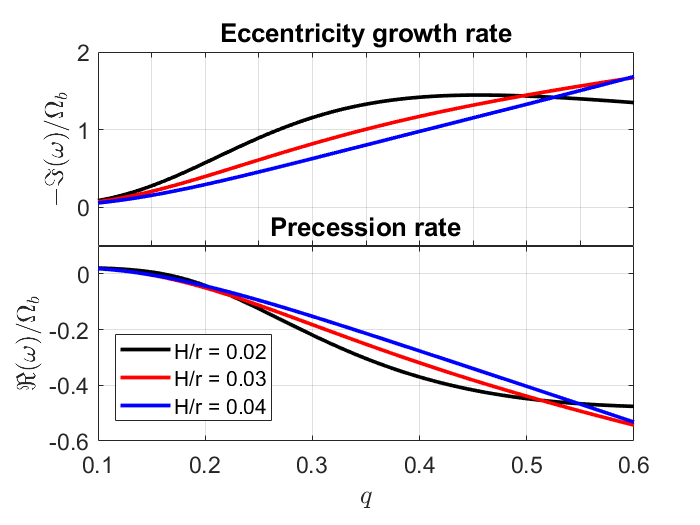}
    \caption{Eccentricity growth rate (upper panel) and the precession rate (lower panel) as a function of the binary mass ratio using the 3D equations. The disk outer radius is $r_{\rm out}=0.9\,r_{\rm l}$.}
    \label{fig:growth}
\end{figure}

Fig.~\ref{fig:growth} shows the eccentricity growth rate and the precession rate as a function of the mass ratio. In all cases, the disk is truncated at $r_{\rm out}=0.9\,r_{\rm l}$ and we solve the 3D equations. We choose the upper limit for the tidal truncation radius which means that we consider the most positive precession rate possible (see Fig.~\ref{fig:rout}). For these parameters, the disk undergoes prograde apsidal precession only for small mass ratios, and large mass ratios undergo retrograde precession. Eccentricity growth for high mass ratio binaries has been seen in 2D simulations. Figure~19 in \cite{Kley2008} shows that the eccentricity growth rate is high up to $q=1.0$, although the direction of the apsidal precession is not discussed. 
Observations show that NSHs are common in systems with large mass ratios and this is in agreement with our results.

We note that there are a few observed systems with high mass ratios that undergo PSHs \citep{Bruch2023}.
We speculate that this may be possible as a result of a surface density profile that is peaked towards the outer edge of the disk. We have only considered power law surface density profiles in our calculations, and we leave exploration of evolving densities to future work.

\section{Conclusions}
\label{sec:concs}

We propose that NSHs in cataclysmic variables may arise from retrograde {\it apsidal} precession of an eccentric disk.  The standard model currently relies on a disk that is tilted with respect to the binary orbital plane. However, there is no satisfactory explanation for how the disk tilts and how it remains tilted with the continuous inflow of material in the binary orbital plane from the companion star. Retrograde apsidal precession offers a natural alternative explanation that avoids these long-standing difficulties.

With linear theory for eccentric disk evolution that includes the effects of the disk pressure, the gravity of the secondary, resonant eccentricity growth, and viscous damping, we have shown that the direction of apsidal precession is highly sensitive 
to both the disk outer truncation radius and the disk aspect ratio, and that pressure 
effects can drive retrograde precession even for relatively cool disks.

For low mass ratio systems, appropriate for SU~Uma dwarf novae, modest changes in disk size and temperature can cause the apsidal precession rate to change direction. During superoutbursts, disk expansion and heating can produce eccentric disks in which the inner regions continue to precess retrogradely while the outer disk precesses progradely, allowing the simultaneous appearance of PSHs and NSHs. The differential precession, or even eccentric disk breaking, may lead to the large dissipation that is observed during superoutbursts. NSHs may persist during quiescence if the disk remains in contact with the resonance, or if the eccentricity damping timescale is long compared to the superoutburst recurrence time. This requires sufficiently small $\alpha$ and $H/r$.

For high mass ratio systems where the 3:1 resonance formally lies outside the disk truncation radius, the finite width of the resonance can still excite eccentricity. In these systems, apsidal precession is generally retrograde for reasonable disk parameters, naturally explaining why NSHs are commonly observed. PSHs are observed in a few systems that have high mass ratios, and we speculate that they may be explained by a density distribution that is peaked towards the outer disk edge.

We encourage future theoretical work and observational studies to remain open to the possibility that NSHs arise from retrograde apsidal disk precession rather than a tilted disk undergoing retrograde nodal disk precession. Although the tilted-disk model successfully explains the observed timing of the NSHs,  retrograde apsidal precession could naturally produce similar photometric periods. This alternative mechanism can potentially explain NSHs while alleviating the need for a disk tilt mechanism, which has remained elusive.

\begin{acknowledgments}
We are very grateful to an anonymous referee for a careful review and useful comments that improved the paper. 
%We thank all the people that have made this AASTeX what it is today.  This
%includes but not limited to Bob Hanisch, Chris Biemesderfer, Lee Brotzman,
%Pierre Landau, Arthur Ogawa, Maxim Markevitch, Alexey Vikhlinin and Amy
%Hendrickson. Also special thanks to David Hogg and Daniel Foreman-Mackey
%for the new {\tt\string modern} style design. Considerable help was provided via bug
%reports and hacks from numerous people including Patricio Cubillos, Alex
%Drlica-Wagner, Sean Lake, Michele Bannister, Peter Williams, Jonathan
%Gagne, Arthur Adams, Nicholas Wogan, Aaron Pearlman, Jeff Mangum, and Mark Durre.
\end{acknowledgments}

\bibliographystyle{aasjournalv7}

\begin{thebibliography}{}
\expandafter\ifx\csname natexlab\endcsname\relax\def\natexlab#1{#1}\fi
\providecommand{\url}[1]{\href{#1}{#1}}
\providecommand{\dodoi}[1]{doi:~\href{http://doi.org/#1}{\nolinkurl{#1}}}
\providecommand{\doeprint}[1]{\href{http://ascl.net/#1}{\nolinkurl{http://ascl.net/#1}}}
\providecommand{\doarXiv}[1]{\href{https://arxiv.org/abs/#1}{\nolinkurl{https://arxiv.org/abs/#1}}}

\bibitem[{A. {Bruch}(2023){Bruch}}]{Bruch2023}
{Bruch}, A. 2023, \bibinfo{title}{{TESS light curves of cataclysmic variables - II - Superhumps in old novae and novalike variables},} \mnras, 519, 352, \dodoi{10.1093/mnras/stac3493}

\bibitem[{P.~P. {Eggleton}(1983){Eggleton}}]{Eggleton1983}
{Eggleton}, P.~P. 1983, \bibinfo{title}{{Aproximations to the radii of Roche lobes.},} \apj, 268, 368, \dodoi{10.1086/160960}

\bibitem[{S. Goodchild \& G. Ogilvie(2006)Goodchild \& Ogilvie}]{GO06}
Goodchild, S., \& Ogilvie, G. 2006, \bibinfo{title}{The dynamics of eccentric accretion discs in superhump systems,} Monthly Notices of the Royal Astronomical Society, 368, 1123, \dodoi{10.1111/j.1365-2966.2006.10197.x}

\bibitem[{J.-M. {Hameury} {et~al.}(1998){Hameury}, {Menou}, {Dubus}, {Lasota}, \& {Hure}}]{Hameury1998}
{Hameury}, J.-M., {Menou}, K., {Dubus}, G., {Lasota}, J.-P., \& {Hure}, J.-M. 1998, \bibinfo{title}{{Accretion disc outbursts: a new version of an old model},} \mnras, 298, 1048, \dodoi{10.1046/j.1365-8711.1998.01773.x}

\bibitem[{M. {Hirose} \& Y. {Osaki}(1990){Hirose} \& {Osaki}}]{Hirose1990}
{Hirose}, M., \& {Osaki}, Y. 1990, \bibinfo{title}{{Hydrodynamic Simulations of Accretion Disks in Cataclysmic Variables: Superhump Phenomenon in SU UMa Stars},} \pasj, 42, 135

\bibitem[{M. {Hirose} \& Y. {Osaki}(1993){Hirose} \& {Osaki}}]{Hirose1993}
{Hirose}, M., \& {Osaki}, Y. 1993, \bibinfo{title}{{Superhump Periods in SU Ursae Majoris Stars: Eigenfrequency of the Eccentric Mode of an Accretion Disk},} \pasj, 45, 595

\bibitem[{S. {Ichikawa} {et~al.}(1993){Ichikawa}, {Hirose}, \& {Osaki}}]{Ichikawa1993}
{Ichikawa}, S., {Hirose}, M., \& {Osaki}, Y. 1993, \bibinfo{title}{{Superoutburst and Superhump Phenomena in SU Ursae Majoris Stars: Enhanced Mass-Transfer Episode or Pure Disk Phenomenon?},} \pasj, 45, 243

\bibitem[{A. {Imada} {et~al.}(2018){Imada}, {Yanagisawa}, \& {Kawai}}]{Imada2018}
{Imada}, A., {Yanagisawa}, K., \& {Kawai}, N. 2018, \bibinfo{title}{{On the colour variations of negative superhumps},} \pasj, 70, L4, \dodoi{10.1093/pasj/psy068}

\bibitem[{L.~M. {Jordan} {et~al.}(2021){Jordan}, {Kley}, {Picogna}, \& {Marzari}}]{Jordan2021}
{Jordan}, L.~M., {Kley}, W., {Picogna}, G., \& {Marzari}, F. 2021, \bibinfo{title}{{Disks in close binary stars. {\ensuremath{\gamma}}-Cephei revisited},} \aap, 654, A54, \dodoi{10.1051/0004-6361/202141248}

\bibitem[{L.~M. {Jordan} {et~al.}(2024){Jordan}, {Wehner}, \& {Kuiper}}]{Jordan2024}
{Jordan}, L.~M., {Wehner}, D., \& {Kuiper}, R. 2024, \bibinfo{title}{{Two-dimensional simulations of disks in close binaries: Simulating outburst cycles in cataclysmic variables},} \aap, 689, A354, \dodoi{10.1051/0004-6361/202348726}

\bibitem[{A. {Joshi} {et~al.}(2025){Joshi}, {Tappert}, {Catelan}, {Schmidtobreick}, \& {Singh}}]{Joshi2025}
{Joshi}, A., {Tappert}, C., {Catelan}, M., {Schmidtobreick}, L., \& {Singh}, M. 2025, \bibinfo{title}{{A tale of three cataclysmic variables with distinct superhumps},} \aap, 702, A70, \dodoi{10.1051/0004-6361/202553810}

\bibitem[{M. {Kimura} \& Y. {Osaki}(2021){Kimura} \& {Osaki}}]{Kimura2021}
{Kimura}, M., \& {Osaki}, Y. 2021, \bibinfo{title}{{KIC 9406652: A laboratory for tilted disks in cataclysmic variable stars. II. Modeling of the orbital light curves},} \pasj, 73, 1225, \dodoi{10.1093/pasj/psab069}

\bibitem[{M. {Kimura} {et~al.}(2020){Kimura}, {Osaki}, \& {Kato}}]{Kimura2020}
{Kimura}, M., {Osaki}, Y., \& {Kato}, T. 2020, \bibinfo{title}{{KIC 9406652: A laboratory for tilted disks in cataclysmic variable stars},} \pasj, 72, 94, \dodoi{10.1093/pasj/psaa088}

\bibitem[{A.~R. {King} {et~al.}(2013){King}, {Livio}, {Lubow}, \& {Pringle}}]{King2013}
{King}, A.~R., {Livio}, M., {Lubow}, S.~H., \& {Pringle}, J.~E. 2013, \bibinfo{title}{{Accretion disc viscosity: what do warped discs tell us?},} \mnras, 431, 2655, \dodoi{10.1093/mnras/stt364}

\bibitem[{W. {Kley} {et~al.}(2008){Kley}, {Papaloizou}, \& {Ogilvie}}]{Kley2008}
{Kley}, W., {Papaloizou}, J.~C.~B., \& {Ogilvie}, G.~I. 2008, \bibinfo{title}{{Simulations of eccentric disks in close binary systems},} \aap, 487, 671, \dodoi{10.1051/0004-6361:200809953}

\bibitem[{J.-P. {Lasota}(2001){Lasota}}]{Lasota2001}
{Lasota}, J.-P. 2001, \bibinfo{title}{{The disc instability model of dwarf novae and low-mass X-ray binary transients},} \nar, 45, 449, \dodoi{10.1016/S1387-6473(01)00112-9}

\bibitem[{S.~H. {Lubow}(1991{\natexlab{a}}){Lubow}}]{Lubow1991b}
{Lubow}, S.~H. 1991{\natexlab{a}}, \bibinfo{title}{{Simulations of Tidally Driven Eccentric Instabilities with Application to Superhumps},} \apj, 381, 268, \dodoi{10.1086/170648}

\bibitem[{S.~H. {Lubow}(1991{\natexlab{b}}){Lubow}}]{Lubow1991}
{Lubow}, S.~H. 1991{\natexlab{b}}, \bibinfo{title}{{A Model for Tidally Driven Eccentric Instabilities in Fluid Disks},} \apj, 381, 259, \dodoi{10.1086/170647}

\bibitem[{S.~H. {Lubow}(1992{\natexlab{a}}){Lubow}}]{Lubow1992}
{Lubow}, S.~H. 1992{\natexlab{a}}, \bibinfo{title}{{Dynamics of Eccentric Disks with Application to Superhump Binaries},} \apj, 401, 317, \dodoi{10.1086/172062}

\bibitem[{S.~H. {Lubow}(1992{\natexlab{b}}){Lubow}}]{Lubow1992a}
{Lubow}, S.~H. 1992{\natexlab{b}}, \bibinfo{title}{{Tidally Driven Inclination Instability in Keplerian Disks},} \apj, 398, 525, \dodoi{10.1086/171877}

\bibitem[{S.~H. Lubow(2010)Lubow}]{Lubow2010}
Lubow, S.~H. 2010, \bibinfo{title}{Eccentricity growth rates of tidally distorted discs,} Monthly Notices of the Royal Astronomical Society, 406, 2777, \dodoi{10.1111/j.1365-2966.2010.16875.x}

\bibitem[{S.~H. {Lubow}(2012){Lubow}}]{Lubow2012}
{Lubow}, S.~H. 2012, \bibinfo{title}{{Erratum: Eccentricity growth rates of tidally distorted discs},} \mnras, 423, 3776, \dodoi{10.1111/j.1365-2966.2012.21217.x}

\bibitem[{N. Meyer-Vernet \& B. Sicardy(1987)Meyer-Vernet \& Sicardy}]{Meyer87}
Meyer-Vernet, N., \& Sicardy, B. 1987, \bibinfo{title}{On the physics of resonant disk-satellite interaction,} Icarus, 69, 157, \dodoi{https://doi.org/10.1016/0019-1035(87)90011-X}

\bibitem[{J.~R. {Murray}(1996){Murray}}]{Murray1996}
{Murray}, J.~R. 1996, \bibinfo{title}{{SPH simulations of tidally unstable accretion discs in cataclysmic variables},} \mnras, 279, 402, \dodoi{10.1093/mnras/279.2.402}

\bibitem[{J.~R. {Murray}(1998){Murray}}]{Murray1998b}
{Murray}, J.~R. 1998, \bibinfo{title}{{Simulations of superhumps and superoutbursts},} \mnras, 297, 323, \dodoi{10.1046/j.1365-8711.1998.01504.x}

\bibitem[{J.~R. {Murray}(2000){Murray}}]{Murray2000}
{Murray}, J.~R. 2000, \bibinfo{title}{{The precession of eccentric discs in close binaries},} \mnras, 314, L1, \dodoi{10.1046/j.1365-8711.2000.03424.x}

\bibitem[{J.~R. {Murray} \& P.~J. {Armitage}(1998){Murray} \& {Armitage}}]{Murray1998}
{Murray}, J.~R., \& {Armitage}, P.~J. 1998, \bibinfo{title}{{Tilted accretion discs in cataclysmic variables: tidal instabilities and superhumps},} \mnras, 300, 561, \dodoi{10.1046/j.1365-8711.1998.01924.x}

\bibitem[{G.~I. {Ogilvie}(2001){Ogilvie}}]{Ogilvie2001}
{Ogilvie}, G.~I. 2001, \bibinfo{title}{{Non-linear fluid dynamics of eccentric discs},} \mnras, 325, 231, \dodoi{10.1046/j.1365-8711.2001.04416.x}

\bibitem[{G.~I. Ogilvie(2007)Ogilvie}]{Ogilvie07}
Ogilvie, G.~I. 2007, \bibinfo{title}{Mean-motion resonances in satellite-disc interactions,} Monthly Notices of the Royal Astronomical Society, 374, 131–149, \dodoi{10.1111/j.1365-2966.2006.11141.x}

\bibitem[{G.~I. {Ogilvie}(2008){Ogilvie}}]{Ogilvie2008}
{Ogilvie}, G.~I. 2008, \bibinfo{title}{{3D eccentric discs around Be stars},} \mnras, 388, 1372, \dodoi{10.1111/j.1365-2966.2008.13484.x}

\bibitem[{G.~I. {Ogilvie} \& A.~J. {Barker}(2014){Ogilvie} \& {Barker}}]{Ogilvie2014}
{Ogilvie}, G.~I., \& {Barker}, A.~J. 2014, \bibinfo{title}{{Local and global dynamics of eccentric astrophysical discs},} \mnras, 445, 2621, \dodoi{10.1093/mnras/stu1795}

\bibitem[{M. {Ohana} {et~al.}(2025){Ohana}, {Jiang}, {Blaes}, \& {Oyang}}]{Ohana2025}
{Ohana}, M., {Jiang}, Y.-F., {Blaes}, O., \& {Oyang}, B. 2025, \bibinfo{title}{{Simulations of Eccentricity Growth in Compact Binary Accretion Disks with Magnetohydrodynamic Turbulence},} \apj, 979, 128, \dodoi{10.3847/1538-4357/ad9ddb}

\bibitem[{T. {Ohshima} {et~al.}(2012){Ohshima}, {Kato}, {Pavlenko}, {Itoh}, {de Miguel}, {Krajci}, {Akazawa}, {Shiokawa}, {Stein}, {Baklanov}, {Samsonov}, {Antonyuk}, {Andreev}, {Imamura}, {Hambsch}, {Maehara}, {Ruiz}, {Nakagawa}, {Kasai}, {Boitnott}, {Virtanen}, \& {Miller}}]{Ohshima2012}
{Ohshima}, T., {Kato}, T., {Pavlenko}, E.~P., {et~al.} 2012, \bibinfo{title}{{Discovery of Negative Superhumps during a Superoutburst of 2011 January in ER Ursae Majoris},} \pasj, 64, L3, \dodoi{10.1093/pasj/64.4.L3}

\bibitem[{T. {Ohshima} {et~al.}(2014){Ohshima}, {Kato}, {Pavlenko}, {Akazawa}, {Imamura}, {Tanabe}, {de Miguel}, {Stein}, {Itoh}, {Hambsch}, {Dubovsky}, {Kudzej}, {Krajci}, {Baklanov}, {Samsonov}, {Antonyuk}, {Malanushenko}, {Andreev}, {Noguchi}, {Ogura}, {Nomoto}, {Ono}, {Nakagawa}, {Taniuchi}, {Aoki}, {Kawabata}, {Kimura}, {Masumoto}, {Kobayashi}, {Matsumoto}, {Shiokawa}, {Shugarov}, {Katysheva}, {Voloshina}, {Zemko}, {Kasai}, {Ruiz}, {Maehara}, {Virnina}, {Virtanen}, {Miller}, {Boitnott}, {Littlefield}, {James}, {Tordai}, {Robert}, {Padovan}, \& {Miyashita}}]{Ohshima2014}
{Ohshima}, T., {Kato}, T., {Pavlenko}, E., {et~al.} 2014, \bibinfo{title}{{Study of negative and positive superhumps in ER Ursae Majoris},} \pasj, 66, 67, \dodoi{10.1093/pasj/psu038}

\bibitem[{A. {Olech} {et~al.}(2009){Olech}, {Rutkowski}, \& {Schwarzenberg-Czerny}}]{Olech2009}
{Olech}, A., {Rutkowski}, A., \& {Schwarzenberg-Czerny}, A. 2009, \bibinfo{title}{{Curious Variables Experiment: SDSS J210014.12+004446.0 - dwarf nova with negative and positive superhumps},} \mnras, 399, 465, \dodoi{10.1111/j.1365-2966.2009.15298.x}

\bibitem[{Y. {Osaki}(1974){Osaki}}]{Osaki1974}
{Osaki}, Y. 1974, \bibinfo{title}{{An Accretion Model for the Outbursts of U Geminorum Stars},} \pasj, 26, 429, \dodoi{10.1093/pasj/26.4.429}

\bibitem[{Y. {Osaki}(1989){Osaki}}]{Osaki1989}
{Osaki}, Y. 1989, \bibinfo{title}{{A Model for the Superoutburst Phenomenon of SU Ursae Majoris Stars},} \pasj, 41, 1005, \dodoi{10.1093/pasj/41.5.1005}

\bibitem[{Y. {Osaki}(1996){Osaki}}]{Osaki1996}
{Osaki}, Y. 1996, \bibinfo{title}{{Dwarf-Nova Outbursts},} \pasp, 108, 39, \dodoi{10.1086/133689}

\bibitem[{Y. {Osaki}(2005){Osaki}}]{Osaki2005}
{Osaki}, Y. 2005, \bibinfo{title}{{The disk instability model for dwarf nova outbursts},} Proceedings of the Japan Academy, Series B, 81, 291, \dodoi{10.2183/pjab.81.291}

\bibitem[{Y. {Osaki} \& T. {Kato}(2013){Osaki} \& {Kato}}]{Osaki2013}
{Osaki}, Y., \& {Kato}, T. 2013, \bibinfo{title}{{The Cause of the Superoutburst in SU UMa Stars is Finally Revealed by Kepler Light Curve of V1504 Cygni},} \pasj, 65, 50, \dodoi{10.1093/pasj/65.3.50}

\bibitem[{Y. {Osaki} \& T. {Kato}(2014){Osaki} \& {Kato}}]{Osaki2014}
{Osaki}, Y., \& {Kato}, T. 2014, \bibinfo{title}{{A further study of superoutbursts and superhumps in SU UMa stars by the Kepler light curves of V1504 Cygni and V344 Lyrae},} \pasj, 66, 15, \dodoi{10.1093/pasj/pst015}

\bibitem[{Y. {Osaki} \& F. {Meyer}(2003){Osaki} \& {Meyer}}]{osaki2003}
{Osaki}, Y., \& {Meyer}, F. 2003, \bibinfo{title}{{Is evidence for enhanced mass transfer during dwarf-nova outbursts well substantiated?},} \aap, 401, 325, \dodoi{10.1051/0004-6361:20030115}

\bibitem[{M. {Overton} {et~al.}(2025){Overton}, {Martin}, {Lubow}, \& {Lepp}}]{Overton2025}
{Overton}, M., {Martin}, R.~G., {Lubow}, S.~H., \& {Lepp}, S. 2025, \bibinfo{title}{{Disc breaking through forced eccentricity growth},} \mnras, 540, L41, \dodoi{10.1093/mnrasl/slaf029}

\bibitem[{B. {Oyang} {et~al.}(2021){Oyang}, {Jiang}, \& {Blaes}}]{Oyang2021}
{Oyang}, B., {Jiang}, Y.-F., \& {Blaes}, O. 2021, \bibinfo{title}{{Investigating lack of accretion disc eccentricity growth in a global 3D MHD simulation of a superhump system},} \mnras, 505, 1, \dodoi{10.1093/mnras/stab1212}

\bibitem[{B. {Paczynski}(1977){Paczynski}}]{Paczynski1977}
{Paczynski}, B. 1977, \bibinfo{title}{{A model of accretion disks in close binaries.},} \apj, 216, 822, \dodoi{10.1086/155526}

\bibitem[{J. {Patterson} {et~al.}(1993){Patterson}, {Thomas}, {Skillman}, \& {Diaz}}]{Patterson1993}
{Patterson}, J., {Thomas}, G., {Skillman}, D.~R., \& {Diaz}, M. 1993, \bibinfo{title}{{The 1991 V603 Aquilae Campaign: Superhumps and P-Dots},} \apjs, 86, 235, \dodoi{10.1086/191777}

\bibitem[{J. {Patterson} {et~al.}(2005){Patterson}, {Kemp}, {Harvey}, {Fried}, {Rea}, {Monard}, {Cook}, {Skillman}, {Vanmunster}, {Bolt}, {Armstrong}, {McCormick}, {Krajci}, {Jensen}, {Gunn}, {Butterworth}, {Foote}, {Bos}, {Masi}, \& {Warhurst}}]{Patterson2005}
{Patterson}, J., {Kemp}, J., {Harvey}, D.~A., {et~al.} 2005, \bibinfo{title}{{Superhumps in Cataclysmic Binaries. XXV. q$_{crit}$, ɛ(q), and Mass-Radius},} \pasp, 117, 1204, \dodoi{10.1086/447771}

\bibitem[{J.~E. {Pringle}(1981){Pringle}}]{Pringle1981}
{Pringle}, J.~E. 1981, \bibinfo{title}{{Accretion discs in astrophysics},} \araa, 19, 137, \dodoi{10.1146/annurev.aa.19.090181.001033}

\bibitem[{N.~I. {Shakura} \& R.~A. {Sunyaev}(1973){Shakura} \& {Sunyaev}}]{Shakura1973}
{Shakura}, N.~I., \& {Sunyaev}, R.~A. 1973, \bibinfo{title}{{Black holes in binary systems. Observational appearance.},} \aap, 24, 337

\bibitem[{J.~C. {Simpson} \& M.~A. {Wood}(1998){Simpson} \& {Wood}}]{Simpson1998}
{Simpson}, J.~C., \& {Wood}, M.~A. 1998, \bibinfo{title}{{Time Series Energy Production in Smoothed Particle Hydrodynamics Accretion Disks: Superhumps in the AM Canum Venaticorum Stars},} \apj, 506, 360, \dodoi{10.1086/306221}

\bibitem[{R.~D. Skeel \& M. Berzins(1990)Skeel \& Berzins}]{Skeel90}
Skeel, R.~D., \& Berzins, M. 1990, \bibinfo{title}{A Method for the Spatial Discretization of Parabolic Equations in One Space Variable,} SIAM J. Sci. Stat. Comput., 11, 1–32

\bibitem[{J. {Smak}(1984){Smak}}]{Smak1984}
{Smak}, J. 1984, \bibinfo{title}{{Accretion in cataclysmic binaries. IV. Accretion disks in dwarf novae.},} \actaa, 34, 161

\bibitem[{J. {Smak}(2009){Smak}}]{Smak2009}
{Smak}, J. 2009, \bibinfo{title}{{On the Origin of Tilted Disks and Negative Superhumps},} \actaa, 59, 419, \dodoi{10.48550/arXiv.0910.2541}

\bibitem[{A.~J. {Smith} {et~al.}(2007){Smith}, {Haswell}, {Murray}, {Truss}, \& {Foulkes}}]{Smith2007}
{Smith}, A.~J., {Haswell}, C.~A., {Murray}, J.~R., {Truss}, M.~R., \& {Foulkes}, S.~B. 2007, \bibinfo{title}{{Comprehensive simulations of superhumps},} \mnras, 378, 785, \dodoi{10.1111/j.1365-2966.2007.11840.x}

\bibitem[{M. {Still} {et~al.}(2010){Still}, {Howell}, {Wood}, {Cannizzo}, \& {Smale}}]{Still2010}
{Still}, M., {Howell}, S.~B., {Wood}, M.~A., {Cannizzo}, J.~K., \& {Smale}, A.~P. 2010, \bibinfo{title}{{Quiescent Superhumps Detected in the Dwarf Nova V344 Lyrae by Kepler},} \apjl, 717, L113, \dodoi{10.1088/2041-8205/717/2/L113}

\bibitem[{Q.-B. {Sun} {et~al.}(2024{\natexlab{a}}){Sun}, {Qian}, {Zhu}, {Li}, {Li}, \& {Li}}]{Sun2024b}
{Sun}, Q.-B., {Qian}, S.-B., {Zhu}, L.-Y., {et~al.} 2024{\natexlab{a}}, \bibinfo{title}{{Tilted Disk Precession and Negative Superhumps in HS 2325+8205: A Multiwindow Analysis},} \apj, 974, 132, \dodoi{10.3847/1538-4357/ad6f05}

\bibitem[{Q.-B. {Sun} {et~al.}(2024{\natexlab{b}}){Sun}, {Qian}, {Zhu}, {Liao}, {Zhao}, {Li}, {Shi}, \& {Li}}]{Sun2024}
{Sun}, Q.-B., {Qian}, S.-B., {Zhu}, L.-Y., {et~al.} 2024{\natexlab{b}}, \bibinfo{title}{{Nine New Cataclysmic Variable Stars with Negative Superhumps},} \apj, 962, 123, \dodoi{10.3847/1538-4357/ad0f1c}

\bibitem[{J. {Teyssandier} \& G.~I. {Ogilvie}(2016){Teyssandier} \& {Ogilvie}}]{Teyssandier2016}
{Teyssandier}, J., \& {Ogilvie}, G.~I. 2016, \bibinfo{title}{{Growth of eccentric modes in disc-planet interactions},} \mnras, 458, 3221, \dodoi{10.1093/mnras/stw521}

\bibitem[{D.~M. {Thomas} \& M.~A. {Wood}(2015){Thomas} \& {Wood}}]{Thomas2015}
{Thomas}, D.~M., \& {Wood}, M.~A. 2015, \bibinfo{title}{{The Emergence of Negative Superhumps in Cataclysmic Variables: Smoothed Particle Hydrodynamics Simulations},} \apj, 803, 55, \dodoi{10.1088/0004-637X/803/2/55}

\bibitem[{B. {Warner}(1995){Warner}}]{Warner1995}
{Warner}, B. 1995, {Cataclysmic variable stars}, Vol.~28

\bibitem[{L. {Wei} \& Q. {Shengbang}(2023){Wei} \& {Shengbang}}]{Liu2023}
{Wei}, L., \& {Shengbang}, Q. 2023, \bibinfo{title}{{Investigation of Superhumps in SU UMa-type Dwarf Novae Based on the Observations of TESS},} \apj, 954, 135, \dodoi{10.3847/1538-4357/acebdf}

\bibitem[{R. {Whitehurst}(1988){Whitehurst}}]{Whitehurst1988}
{Whitehurst}, R. 1988, \bibinfo{title}{{Numerical simulations of accretion discs - I. Superhumps : a tidal phenomenon of accretion discs.},} \mnras, 232, 35, \dodoi{10.1093/mnras/232.1.35}

\bibitem[{R. {Whitehurst} \& A. {King}(1991){Whitehurst} \& {King}}]{Whitehurst1991}
{Whitehurst}, R., \& {King}, A. 1991, \bibinfo{title}{{Superhumps, resonances and accretion discs.},} \mnras, 249, 25, \dodoi{10.1093/mnras/249.1.25}

\bibitem[{M.~A. {Wood} {et~al.}(2009){Wood}, {Thomas}, \& {Simpson}}]{Wood2009}
{Wood}, M.~A., {Thomas}, D.~M., \& {Simpson}, J.~C. 2009, \bibinfo{title}{{SPH simulations of negative (nodal) superhumps: a parametric study},} \mnras, 398, 2110, \dodoi{10.1111/j.1365-2966.2009.15252.x}

\end{thebibliography}



\end{document}